\documentclass[aps,prb,groupedaddress,twocolumn,amsmath]{revtex4}
\usepackage{dcolumn}
    \usepackage{bm}
    \usepackage{graphicx,amssymb}
\usepackage{bm}
\usepackage{amsmath}
\newcommand\beq{\begin{equation}}
\newcommand\eeq{\end{equation}}
\newcommand\beqa{\begin{eqnarray}}
\newcommand\eeqa{\end{eqnarray}}
\newcommand{\nn}{\nonumber\\}

\begin{document} 
\title{Molecular dynamics and theory for the contact values of the radial
       distribution functions of hard-disk fluid mixtures}
\author{Stefan Luding}
 \email[]{s.luding@tnw.tudelft.nl}
\homepage[]{http://www.dct.tudelft.nl/part/welcomePTG.html}
\affiliation{Particle Technology, DelftChemTech, 
             Julianalaan 136, 2628 BL Delft, The Netherlands}
\author{Andr\'es Santos}
\email[]{andres@unex.es}
\homepage[]{http://www.unex.es/fisteor/andres/}
\affiliation{Departamento de F\'{\i}sica, Universidad de Extremadura, 
             Badajoz, E-06071, Spain}
\date{\today}

\begin{abstract} 
We report molecular dynamics results for the contact values of the radial 
distribution functions of binary additive mixtures of hard disks. 
The simulation data are compared with theoretical predictions from 
expressions proposed by Jenkins and Mancini 
[J. Appl. Mech. \textbf{54}, 27 ({1987})] 
and Santos et al.\ 
[J. Chem. Phys. \textbf{117}, 5785 (2002)]. 
Both theories agree quantitatively within a very small margin,
which renders the former still a very useful and simple tool to 
work with. 
The latter (higher-order and self-consistent) theory provides
a small qualitative correction for low densities and is superior 
especially in the high-density domain.
\end{abstract}

\maketitle

\section{Introduction\label{sec1}}

Model systems of hard disks and hard spheres are useful for the derivation 
of rigorous results in statistical mechanics as well as
in perturbation theories of fluids.\cite{HM86} Hard disks and hard 
spheres are also relevant for the modeling of mesoscopic systems 
such as colloidal suspensions\cite{L94} and granular matter.\cite{C90}
Apart from its academic interest, the study of two-dimensional systems 
is important in the context of monolayer adsorption on solid 
surfaces.\cite{S76} Recently, the equation of state of hard disks has 
been \textit{experimentally} measured in charge-stabilized 
colloidal particles suspended in water and confined by a laser beam.\cite{BBHG03}
While most of the studies are restricted to monodisperse fluids, it 
is obviously  important to consider the \textit{polydisperse} 
character of the system, especially in applications to mesoscopic matter. 
The equation of state, as well as nonequilibrium  transport properties,  
of bidisperse systems of inelastic hard disks {have been  discussed
in the literature},
see Refs.\ 
\onlinecite{JM87,willits99,alam02,alam02b,alam03,garzo03,garzo03b,MG03,rahaman03,dahl03,dahl04}
and references therein.


The state of an additive $m$-component fluid mixture of $N=\sum_{i=1}^{m} N_i$ 
hard disks is characterized by the total number 
density $\rho$, the set of mole fractions $\textbf{x}\equiv\{x_1,x_2,\ldots,x_m\}$, 
with $x_i=N_i/N$, and the set of diameters 
$\bm{\sigma}\equiv\{\sigma_1,\sigma_2,\ldots,\sigma_m\}$. Instead of $\rho$, 
the area fraction 
$\nu=(\pi/4)\rho\langle \sigma^2\rangle$, where $\langle \sigma^n\rangle\equiv 
\sum_i x_i\sigma_i^n$ is the $n$-th moment of the 
size distribution, can be used to characterize the density of the system.
The spatial correlation between two disks of species $i$ and $j$ separated 
by a distance $r$ is measured by the radial distribution 
function (RDF) $g_{ij}(r;\nu,\textbf{x},\bm{\sigma})$. The contact values 
\beq
\chi_{ij}(\nu,\textbf{x},\bm{\sigma})\equiv g_{ij}(\sigma_{ij};\nu,\textbf{x},
\bm{\sigma}), \quad \sigma_{ij}\equiv 
(\sigma_i+\sigma_j)/2,
\label{1.1}
\eeq
of the RDFs are of special interest since they appear in the Enskog kinetic 
theory of dense fluids\cite{FK72} and, more important, 
they are directly related to the equation of state (EOS) of the fluid via the 
virial theorem,\cite{RS82}
\beqa
Z (\nu,\textbf{x},\bm{\sigma})
   & \equiv & \frac{p}{\rho k_B T} \nn
   & = & 1 + 2\nu \sum_{i,j}x_ix_j
                  \frac{\sigma_{ij}^2}
                       {\langle\sigma^2\rangle}
              \chi_{ij}(\nu,\textbf{x},\bm{\sigma}).
\label{1.2}
\eeqa
Alternatively, the compressibility factor $Z$ can be expressed as\cite{NHSC98}
\beq
Z_{w} (\nu,\textbf{x},\bm{\sigma})=\sum_i x_i \chi_{iw}(\nu,\textbf{x},\bm{\sigma}),
\label{1.3}
\eeq
where $\chi_{iw}(\nu,\textbf{x},\bm{\sigma})$ denotes the density of species 
$i$ at contact with a planar hard wall, relative to the 
associated bulk density. Its expression  can be obtained from that of 
$\chi_{ij}(\nu,\textbf{x},\bm{\sigma})$ by assuming the 
wall to be a component of the mixture present in zero concentration and having 
an infinite diameter:\cite{NHSC98,HAB76}
\beq
\chi_{iw}(\nu,\textbf{x},\bm{\sigma})=\lim_{\sigma_j\to\infty} \, \, 
         \lim_{x_j\to 0} \, \chi_{ij}(\nu,\textbf{x},\bm{\sigma}).
\label{1.4}
\eeq
The subscript $w$ in $Z_w$ has been used to emphasize that Eq.\ (\ref{1.3}) 
represents a route alternative to Eq.\ (\ref{1.2}) to 
get the EOS of the hard-disk polydisperse fluid. Of course, $Z=Z_w$ in an exact 
description, but $Z$ and $Z_w$ may differ when 
dealing with approximations. Thus a stringent consistency condition for an 
approximate theory of $\chi_{ij}$ is to  yield the same 
EOS through both routes.

The aim of this paper is two-fold. First, we present (new, accurate) 
molecular dynamics results for $\chi_{ij}$ in the case of a bidisperse 
hard-disk fluid mixture with a size ratio $\sigma_1/\sigma_2=1/2$. 
Next, those simulation data are compared against theoretical 
predictions by Jenkins and Mancini\cite{JM87} and by 
Santos et al.\cite{SYH02}
As will be seen, both theories agree quantitatively 
well with the simulation data but the latter is slightly superior in the 
high-density fluid regime ($0.3\lesssim \nu\lesssim 0.7$). 
The theoretical proposals for $\chi_{ij}$ are presented in Sec.\ \ref{sec1bis} 
and the simulation method is outlined in Sec.\ \ref{sec2}. 
The results are presented  and discussed in Sec.\ \ref{sec3} 
and we close the paper in Sec.\ \ref{sec4} with some final remarks.

\section{Theoretical approximations}

\label{sec1bis}

In this section, different approximations for the contact values of 
the radial distribution function are reviewed, first for the monodisperse
and then for the polydisperse case.  

\subsection{The monodisperse case}

\begin{figure}[htbp]
\includegraphics[clip=,width=0.90 \columnwidth]{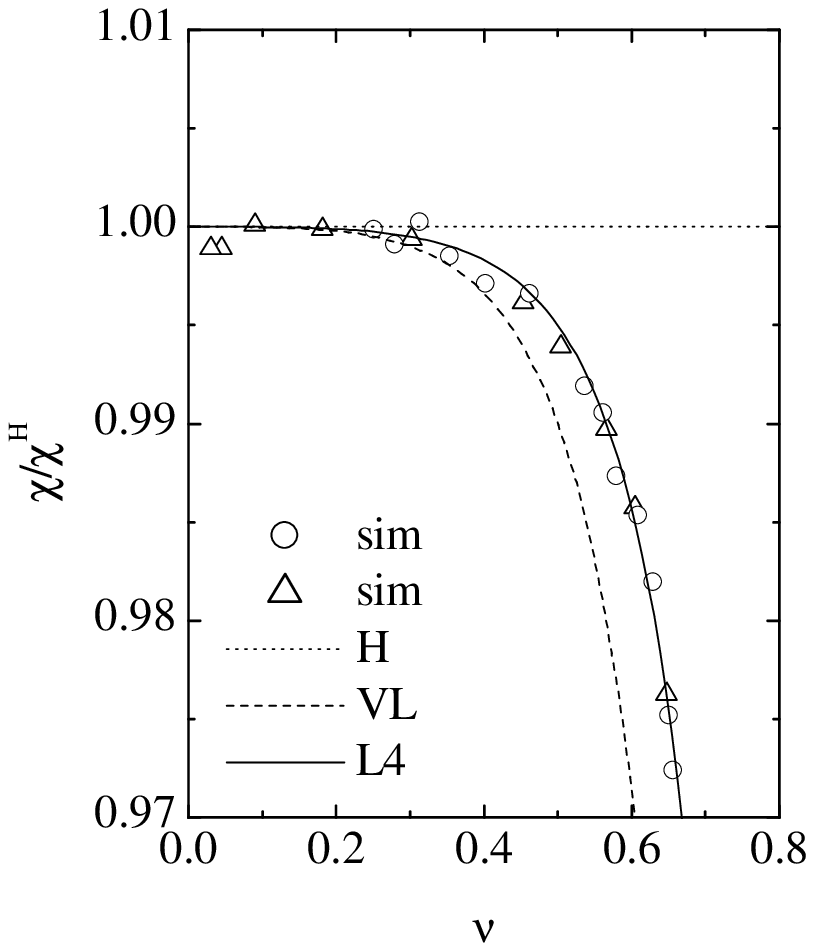}
\caption{Density dependence of the ratio $\chi/\chi^{\text{H}}$, where
unity (dotted line) corresponds to Eq.\ (\protect\ref{1.7}). 
The circles\protect\cite{Lu01,Lu02,Lu04} and triangles\protect\cite{EL85} 
are simulation data, while the dashed and solid lines are the theoretical 
predictions given by Eqs.\ (\protect\ref{1.8}) and (\protect\ref{1.9}),
respectively. 
\label{gmono}}
\end{figure}

For the sake of completeness,  let us begin with the one-component fluid 
before considering the more general polydisperse fluid.
Perhaps the most widely used approximation for the contact value $\chi(\nu)$ 
of the RDF of the monodisperse hard-disk fluid
is the one proposed by Henderson in 1975:\cite{H75}
\beq
\chi^{\text{H}}(\nu)=\frac{1-7\nu/16}{(1-\nu)^2}.
\label{1.7}
\eeq
Despite the simplicity of this prescription, it provides fairly accurate 
values. 
On the other hand, Eq.\ (\ref{1.7}) tends to overestimate 
the value of $\chi(\nu)$ for high densities of the stable fluid 
phase.\cite{VL82,SHY95,MCG97} This led Verlet and Levesque\cite{VL82} 
to propose the correction
\beq
\chi^{\text{VL}}(\nu)=\chi^{\text{H}}(\nu)-\frac{\nu^3}{2^6(1-\nu)^4}.
\label{1.8}
\eeq
A more accurate prescription has recently been proposed by 
one of us:\cite{Lu01,Lu01b,Lu02,Lu04}
\beqa
\chi^{\text{L4}}(\nu)
    &=&\frac{1}{2}
       \left[\chi^{\text{H}}(\nu)+\chi^{\text{VL}}(\nu)\right]\nn
    &=&\chi^{\text{H}}(\nu)-\frac{\nu^3}{2^7(1-\nu)^4}.
\label{1.9}
\eeqa
This is confirmed by Fig.\ \ref{gmono}, where simulation data of 
$\chi(\nu)$,\cite{EL85,Lu01,Lu02,Lu04} relative to Henderson's 
approximation $\chi^{\text{H}}(\nu)$, are compared with the ratios 
$\chi^{\text{VL}}(\nu)/\chi^{\text{H}}(\nu)$ and
$\chi^{\text{L4}}(\nu)/\chi^{\text{H}}(\nu)$. 
We observe that $\chi^{\text{H}}(\nu)$ behaves satisfactorily up 
to an area fraction $\nu\approx 0.3$. 
However, as the density increases and approaches the limit of 
stability of the hard-disk fluid ($\nu\simeq 0.7$),\cite{J99,Lu01,Lu01b,Lu04,Lu02} 
$\chi^{\text{H}}(\nu)$ overestimates the simulation data {(by a 
few percent at $\nu=0.68$)}, while Eq.\ (\ref{1.9}) presents an 
excellent agreement with computer simulations for
densities $\nu \le 0.68$. 

\subsection{The polydisperse case}

In this subsection, the classical result for the RDF contact
value\cite{JM87} is confronted to {more recent findings.\cite{SYH02}}
The derivation is detailed for bulk- and wall-EOS in both cases,
and the agreement/disagreement of the two approaches is discussed.

\subsubsection{Jenkins and Mancini's approximation}

In the case of a hard-disk mixture, a useful approximation for the contact 
values $\chi_{ij}$  was proposed by Jenkins and Mancini in 
1987.\cite{JM87} It reads
\beq
\chi_{ij}^{\text{JM}}(\nu,\textbf{x},\bm{\sigma})
  =\frac{1}{1-\nu}+\frac{9}{16}\frac{\nu}{(1-\nu)^2}z_{ij}
                                 (\textbf{x},\bm{\sigma}),
\label{1.5}
\eeq
where the parameter
\beq
z_{ij}(\textbf{x},\bm{\sigma})
   \equiv \frac{\sigma_{i}\sigma_{j}}{\sigma _{ij}}
          \frac{\langle \sigma\rangle}{\langle \sigma^{2}\rangle}
\label{1.6}
\eeq
contains the whole dependence of $\chi_{ij}$ on the size composition 
through the first two moments.
It is worth mentioning that Eq.\ (\ref{1.5}) was originally proposed 
in the context of \textit{inelastic} disks and thus it is much better
 known by researchers in granular matter theory than by researchers in 
liquid theory. As a matter of fact, Eq.\ (\ref{1.5}) has recently
 been rediscovered by liquid theorists.\cite{BS01}

In the special case where all the disks have the same diameter 
($\sigma_i\to\sigma$), one has $z_{ij}\to 1$ and so
Eq.\ (\ref{1.5}) reduces to Henderson's approximation (\ref{1.7}) 
for monodisperse disks. Thus,  Jenkins and Mancini's approximation 
(\ref{1.5}) represents a simple, straightforward
extension of $\chi^{\text{H}}$ to the polydisperse case. 
As a consequence,  it inherits, by construction, the 
limitations of Henderson's equation (\ref{1.7}) for high 
densities (cf.\ Fig.\ \ref{gmono}).  Moreover, Eq.\ (\ref{1.5}) does not 
yield the same 
EOS through routes (\ref{1.2}) and (\ref{1.3}). Let us first consider 
the standard route (\ref{1.2}). Taking into account the 
mathematical identities (for an arbitrary number of components)
\beq
\sum_{i,j}x_ix_j\sigma_{ij}^2=\frac{\langle\sigma^2\rangle+
\langle\sigma\rangle^2}{2},
\label{n1}
\eeq
\beqa
\sum_{i,j}x_ix_j\sigma_{ij}^2z_{ij}&=&\frac{\langle\sigma\rangle}
{2\langle\sigma^2\rangle}\sum_{i,j}x_ix_j\sigma_i\sigma_j
(\sigma_i+\sigma_j)\nn 
&=& \langle\sigma\rangle^2,
\label{n2}
\eeqa
insertion of Eq.\ (\ref{1.5}) into Eq.\ (\ref{1.2}) yields
\beqa
Z^{\text{JM}}(\nu,\textbf{x},\bm{\sigma})
 & = & 1 + \frac{\nu}{1-\nu}+\mathcal{A}\nu 
                   \frac{1+\nu/8}{(1-\nu)^2} \nn
 & = & 1 + 2 \nu \left [ \frac{1-{\cal A}}{2(1-\nu)}
                         + {\cal A} \chi^{\text{H}}(\nu)
                 \right ] ~,
\label{1.10}
\eeqa
where ${\cal A}\equiv{\langle\sigma\rangle^2} / {\langle\sigma^2\rangle}$
was used as a convenient abbreviation.\cite{Lu01b,Lu02} 
To explore the alternative route (\ref{1.3}), let us
take the limits indicated in Eq.\ (\ref{1.4}) on Eq.\ (\ref{1.5}) 
to get
\beq
\chi_{iw}^{\text{JM}}(\nu,\textbf{x},\bm{\sigma})=
  \frac{1}{1-\nu}+\frac{9}{16}\frac{\nu}{(1-\nu)^2} \,
                      z_{iw}(\textbf{x},\bm{\sigma}),
\label{1.10bis}
\eeq
where
\beq
z_{iw}(\textbf{x},\bm{\sigma})\equiv 2\sigma_{i}\frac
{\langle \sigma\rangle}{\langle \sigma^{2}\rangle}. 
\label{n3}
\eeq
Inserting Eq.\ (\ref{1.10bis}) into Eq.\ (\ref{1.3}) and making use of
\beq
\sum_i x_i z_{iw}=2\mathcal{A},
\label{n5}
\eeq
 one has
\begin{eqnarray}
Z_{w}^{\text{JM}}(\nu,\textbf{x},\bm{\sigma})
 & = & \sum_i x_i \chi_{iw}^{\text{JM}}(\nu,\textbf{x},\bm{\sigma}) \nn 
 & = & Z^{\text{JM}}(\nu,\textbf{x},\bm{\sigma})
     +\mathcal{A} 
                                       \frac{\nu/8}{(1-\nu)} ~.
\label{1.11}
\end{eqnarray}
The inconsistency $Z^{\text{JM}}\neq Z_{w}^{\text{JM}}$ 
appears already to first order in $\nu$.
Comparison with simulation results shows that $Z^{\text{JM}}$ is 
clearly superior to $Z_{w}^{\text{JM}}$.

\subsubsection{Santos, Yuste, and L\'opez de Haro's approximation}

The two limitations of $\chi_{ij}^{\text{JM}}$ just mentioned, 
namely the slavery to Henderson's equation $\chi^{\text{H}}$ and 
the failure to give a common EOS through 
Eqs.\ (\ref{1.2}) and (\ref{1.3}), are remedied by a recent proposal made by 
Santos et al.\cite{SYH02} It reads
\beqa
\label{1.12} 
\chi_{ij}^{\text{SYH}}(\nu,\textbf{x},\bm{\sigma})
  & = & \frac{1}{1-\nu } \nn
  &  &+ \left[ (2-\nu)\chi(\nu)-\frac{2-\nu/2}{1-\nu}\right] 
          z_{ij}(\textbf{x},\bm{\sigma}) \nn
  &  &+ \left[\frac{1-\nu/2}{1-\nu }- (1-\nu)\chi(\nu)\right]
          z_{ij}^2(\textbf{x},\bm{\sigma}) ~, \nn
&& 
\eeqa
where $z_{ij}$ is again given by Eq.\ (\ref{1.6}) and the contact value 
$\chi(\nu)$ of the 
monodisperse fluid can be freely chosen.
Obviously, in the trivial case where all the disks 
have the same size, $z_{ij}=1$ and so 
$\chi_{ij}^{\text{SYH}}(\nu)=\chi(\nu)$. 
Insertion of Eq.\ (\ref{1.12}) 
into Eq.\ (\ref{1.2}) yields the following simple  form:
\beqa
Z^{\text{SYH}}(\nu,\textbf{x},\bm{\sigma})
 & = & 1+\frac{\nu}{1-\nu}
        +\mathcal{A} \nu
         \left[2\chi(\nu)-\frac{1}{1-\nu}\right] \nn
 & = & 1+2 \nu \left [ \frac{1-{\cal A}}{2(1-\nu)}
                        + {\cal A} \chi(\nu) 
               \right ] ~,
\label{1.13}
\eeqa
where use has been made of Eqs.\ (\ref{n1}), (\ref{n2}), and
\beqa
\sum_{i,j}x_ix_j\sigma_{ij}^2z_{ij}^2&=&\frac{\langle\sigma\rangle^2}
{\langle\sigma^2\rangle^2}
\sum_{i,j}x_ix_j\sigma_i^2\sigma_j^2\nn 
&=& \langle\sigma\rangle^2,
\label{n4}
\eeqa
valid again for any number of components. Note the identical form of the second lines in both
Eq.\ (\ref{1.10}) and Eq.\ (\ref{1.13}).
The expression for $\chi_{iw}^{\text{SYH}}$ is given by Eq.\ (\ref{1.12}), 
except that $z_{ij}$ must be replaced by $z_{iw}$. 
When $\chi_{iw}^{\text{SYH}}$ is inserted into Eq.\ (\ref{1.3}) and use 
is made of Eq.\ (\ref{n5}) and of
\beq
\sum_i x_i z_{iw}^2=4\mathcal{A},
\label{n6}
\eeq
it turns out that the EOS (\ref{1.13}) is consistently 
reobtained, i.e.\ $Z^{\text{SYH}}_w \equiv Z^{\text{SYH}}$.

So far, the monodisperse quantity $\chi(\nu)$ remains arbitrary. 
{}From that point of view, Eq.\ (\ref{1.12}) represents a consistent
class of approximations, with free $\chi(\nu)$, rather than a specific 
approximation. 

\subsubsection{Some comments}

It is worth mentioning that Eqs.\ (\ref{1.5}) and (\ref{1.12}) share 
the property that, at a given packing fraction $\nu$,  the whole
dependence of $\chi_{ij}$ on the composition $(\textbf{x},\bm{\sigma})$ 
of the mixture appears through the parameter $z_{ij}$ only. To clarify 
the implications of this, let us  consider two  mixtures 
M and M' having the same packing fraction $\nu$ but strongly differing 
in the set of mole fractions, the sizes of the particles, 
and even the number of components, i.e.\ $(\textbf{x},\bm{\sigma})\neq 
(\textbf{x}',\bm{\sigma}')$. Suppose now that there 
exists  a pair $ij$ in mixture M and another pair $i'j'$ in mixture M' 
such that $z_{ij}(\textbf{x},\bm{\sigma}) = 
z_{i'j'}(\textbf{x}',\bm{\sigma}')$. Then, according to Eqs.\ (\ref{1.5}) 
and (\ref{1.12}), the contact value of the RDF for the pair  
$ij$ in mixture M is the same as that for the pair $i'j'$ in mixture M'. 
This sort of ``universality'' ansatz, which is more general than  
Eqs.\ (\ref{1.5}) or (\ref{1.12}) and is shared by other well-known
proposals for $\chi_{ij}$ of hard-sphere fluids (e.g. the scaled-particle-theory,
Percus--Yevick, and Boubl\'{\i}k--Grundke--Henderson--Lee--Levesque
approximations),\cite{SYH02} 
is of course only approximate. 
However, its enforcement leads to the construction of simple and 
accurate proposals for $\chi_{ij}$ with the help of only a few 
requirements.\cite{SYH02,SYH99} 

Interestingly enough, the EOS (\ref{1.10}) and
(\ref{1.13}) are identical when $\chi=\chi^{\text{H}}$ is used in the 
latter, even though the contact values  $\chi_{ij}$  
used in the derivation are different. More specifically, if 
$\chi=\chi^{\text{H}}$ is used in Eq.\ (\ref{1.12}), then
\beq
\chi_{ij}^{\text{SYH}}-\chi_{ij}^{\text{JM}}
=\frac{\nu/16}{1-\nu}z_{ij}\left(1-z_{ij}\right).
\label{n7}
\eeq
The property $Z^{\text{SYH}}=Z^{\text{JM}}$ is just a consequence 
of $\sum_{i,j}x_ix_j\sigma_{ij}^2z_{ij}=
\sum_{i,j}x_ix_j\sigma_{ij}^2z_{ij}^2$, as follows from Eqs.\ 
(\ref{n2}) and (\ref{n4}).

\begin{figure}[htbp]
\includegraphics[clip=,width=0.90 \columnwidth]{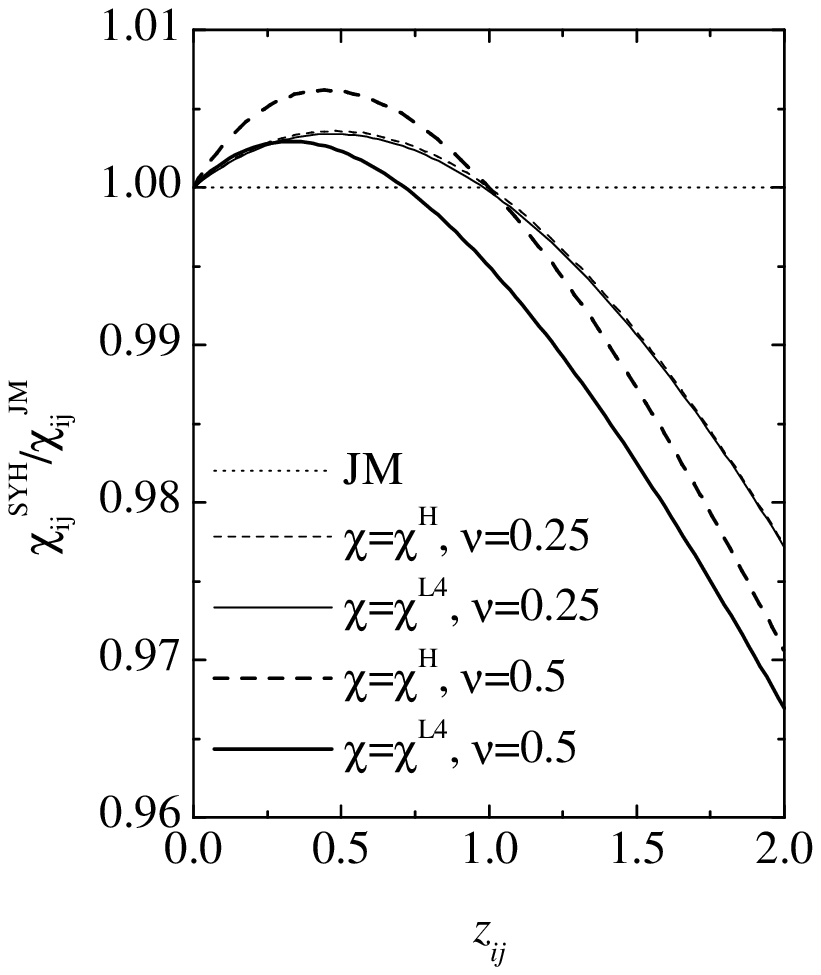}
\caption{Plot of the ratio $\chi_{ij}^{\text{SYH}}/\chi_{ij}^{\text{JM}}$ 
against the parameter $z_{ij}$ for $\nu=0.25$ (thin lines) and 
$\nu=0.50$ (thick lines). The dashed lines correspond to the choice 
$\chi(\nu)=\chi^{\text{H}}(\nu)$ in Eq.\ (\protect\ref{1.12}), 
while the solid lines correspond  to the choice 
$\chi(\nu)=\chi^{\text{L4}}(\nu)$. The dotted line represents the 
JM expression (\protect\ref{1.5}).
\label{syh}}
\end{figure}

If one chooses $\chi(\nu)=\chi^{\text{H}}(\nu)$ then Eq.\ (\ref{1.12}), 
being consistent with the  condition $Z=Z_w$, can be 
expected to become more accurate than Eq.\ (\ref{1.5}), especially for 
highly asymmetric mixtures. Since $\chi^{\text{H}}(\nu)$ is 
fairly good for $\nu\lesssim 0.3$, as Fig.\ \ref{gmono} shows, the 
main difference between Eqs.\ (\ref{1.5}) and (\ref{1.12}) in that 
density domain lies in the functional relation on the parameter $z_{ij}$: 
linear in the case of Eq.\ (\ref{1.5}), quadratic in the case of 
Eq.\ (\ref{1.12}). As a consequence,
\beqa
\chi_{ij}^{\text{SYH}}/\chi_{ij}^{\text{JM}}&>& 1\quad \text{if } 
z_{ij}<1,\nn
&=&1\quad\text{if }z_{ij}=1,\nn
&<&1\quad\text{if }z_{ij}>1,
\label{1.14}
\eeqa
where we emphasize that (\ref{1.14}) refers to $0<\nu\lesssim 0.3$. 
This is illustrated in Fig.\ \ref{syh}, where the ratio 
 $\chi_{ij}^{\text{SYH}}/\chi_{ij}^{\text{JM}}$ is plotted versus 
$z_{ij}$ for $\nu=0.25$ and $\nu=0.5$. In the former case, 
since $\chi^{\text{H}}$ and $\chi^{\text{L4}}$ yield practically 
the same value, the associated two functions given by 
Eq.\ (\ref{1.12}) are hardly distinguishable. On the other hand, 
those functions exhibit visible small differences at $\nu=0.5$ due to 
the fact that $\chi^{\text{H}}$ deviates from $\chi^{\text{L4}}$ by about 
$0.5\%$ (cf.\ Fig.\ \ref{gmono}).

In the spirit of the approximations (\ref{1.12}) and (\ref{1.13}), the 
more accurate the monodisperse function $\chi(\nu)$ the better. 
Therefore, more successful predictions for $\chi_{ij}$ and $Z$ can be 
expected if one chooses for $\chi(\nu)$ an expression more 
refined than Henderson's, such as Eq.\ (\ref{1.9}),  especially for 
high densities. 
In Sec.\ \ref{sec3} we will check all these expectations by comparing 
molecular dynamics results for $\chi_{ij}$ against  
Eqs.\ (\ref{1.5}) and (\ref{1.12}), the latter being complemented by 
the monodisperse prescriptions (\ref{1.7}) and (\ref{1.9}). 

\section{Simulation method\label{sec2}}

Since we are interested in the behavior of rigid particles,
we use an event-driven (ED) method that discretizes the sequence 
of events with {variable time steps for all particles between
collisions, as adapted to the problem.  
This is different from classical molecular dynamics simulations,
where the time step is usually fixed for the numerical integration
of the equations of motion.}

\subsection{Collision Model}

The particles are assumed to be perfectly rigid and follow an
undisturbed motion until a collision occurs as described below.
A change in velocity can occur only at a collision, and
due to their rigidity, the disks collide instantaneously. 
The standard interaction model for instantaneous collisions of
particles with diameters $\sigma_i$ and mass $m_i$ is used in the 
following.
The post-collisional velocities $(\mathbf{v}_i',\mathbf{v}_j')$ of 
two collision partners $ij$
 are given, in terms of
the pre-collisional velocities $(\mathbf{v}_i,\mathbf{v}_j)$, by
\begin{equation}
\mathbf{v}_{i,j}' = \mathbf{v}_{i,j} \mp \frac{2~\mu_{ij}}{m_{i,j}} 
\mathbf{v}_n ~,
\label{eq:collrule}
\end{equation}
where $\mu_{ij} = m_i m_j / (m_i+m_j)$ is the reduced mass and  
$\mathbf{v}_n \equiv \left [ (\mathbf{v}_i - \mathbf{v}_j) 
                         \cdot {\mathbf{n}} \right ] {\mathbf{n}}$ is
the component of the relative velocity $\mathbf{v}_i-\mathbf{v}_j$ parallel to 
the unit vector ${\mathbf{n}}$ pointing along the line connecting the
centers of the colliding particles.
If two particles collide, their velocities are changed according
to Eq.\ (\ref{eq:collrule}).

\subsection{Algorithm}

In the ED simulations, the particles follow an undisturbed translational
motion until an event occurs. An event is either the collision
of two particles or the collision of one particle with a boundary
of a cell (in the linked-cell structure).\cite{allen87}
The cells have no effect on the particle motion here;
they were solely introduced to accelerate the search for future
collision partners in the algorithm.

Simple ED algorithms update the whole system after each event, a
method which is straightforward, but inefficient for large numbers of
particles. In Ref.\ \onlinecite{lubachevsky91} an ED algorithm was introduced
which updates only those two particles involved in the last
collision. 
For the algorithm, a double buffering data structure is implemented,
which contains the `old' status and the `new' status, each consisting
of: time of event, positions, velocities, and event partners. When a
collision occurs, the `old' and `new' status of the participating
particles are exchanged. Thus, the former `new' status becomes the
actual `old' one, while the former `old' status becomes the `new' one
and is then free for the calculation and storage of possible future events.
This seemingly complicated
exchange of information is carried out extremely simply and fast by
only exchanging the pointers to the `new' and `old' status
respectively. Note that the `old' status of particle $i$ has to be kept in
memory, in order to update the time of the next contact, $t_{ij}$,
of particle $i$ with any other object $j$, if the latter, independently,
changed its status due to a collision with yet another particle.
During the simulation such updates
may be necessary several times so that the predicted `new' status
has to be modified.

The minimum of all $t_{ij}$ is stored in the `new' status of particle $i$,
together with the corresponding partner $j$. Depending on the
implementation, positions and velocities after the collision can also
be calculated.  This would be a waste of computer time, since before
the time $t_{ij}$, the predicted partners $i$ and $j$ might be
involved in several collisions with other particles, so that we
apply a delayed update scheme.\cite{lubachevsky91} The minimum
times of event, i.e.\ the times which indicate the next event for a
certain particle, are stored in an ordered heap tree, such that the
next event is found at the top of the heap with a computational effort
of ${O}(1)$; changing the position of one particle in the tree
from the top to a new position needs ${O}(\log N)$ operations.
The search for possible collision partners is accelerated
by the use of a standard linked-cell data structure and consumes ${O}(1)$
of numerical resources per particle. In total, this results in a numerical
effort of ${O}(N \log N)$ for $N$ particles. For a detailed description
of the algorithm see Ref.\ \onlinecite{lubachevsky91}.

\subsection{Computation of $\chi_{ij}$}

The results {for the RDF contact values
are computed indirectly as}
\beq
\chi_{ij} = \frac{\tau_{ij}^{-1} A}
                 {2 \sigma_{ij} N_j \sqrt{\pi T/(2 \mu_{ij})}} ~,
\eeq
where $\tau_{ij}^{-1}$ is the average number of $(ij)$ collisions per 
unit time and per particle of species $i$, $A$ is the area of the system, 
and $T=T_{i,j}=E^{i,j}_{\rm kin}/N_{i,j}$ is the temperature
based on the kinetic energy per particle per degree of freedom.
Note that $\tau_{ij}^{-1}$ is proportional to $N_j$ and hence 
$\chi_{ij}=\chi_{ji}$.

The averages are taken over a few hundred thousand (low density)
up to several millions (high density) collisions per particle,
where the first 20--30\% of the simulation time is typically disregarded, 
so that the average is taken in a reasonably equilibrated state.

\section{Results and discussion\label{sec3}}

We have considered two hard-disk binary mixtures
with a mole fraction of small disks  $x_1$
and a mole fraction of large disks  $x_2$, as
summarized in Table \ref{tab:simpara}.
The diameter ratio of small to large disks is $\sigma_1/\sigma_2={1}/{2}$
in both cases. The corresponding values of the parameters $z_{ij}$ defined by
Eq.\ (\ref{1.6}) are also included in Table \ref{tab:simpara}.
\begin{table}[htb]
\begin{ruledtabular}
\begin{tabular}{lcccccccc}
Mixture & $N_1$ & $N_2$ & $x_1$ & $x_2$ & $\sigma_1/\sigma_2$
                              & $z_{11}$ & $z_{12}$ & $z_{22}$ \\
\hline
A & 450   & 126   & 0.781 & 0.219 & $1/2$  & 0.736  & 0.981 & 1.472 \\
B & 7803  & 1998  & 0.796 & 0.204 & $1/2$  & 0.747  & 0.996 & 1.494 
\end{tabular}
\caption{Simulation parameters for mixtures A and B.}
\label{tab:simpara}
\end{ruledtabular}
\end{table}

Mixtures A and B have nearly the same composition ($x_1\simeq 0.8$, 
$x_2\simeq 0.2$), but the number of particles in mixture B is 
about 17 times larger than in mixture A and so the statistics is better
in case B.  The data of $\chi_{ij}$ for several area fractions
in the interval $0.01 \leq\nu\leq 0.75$ are given in Table \ref{table1}. 
The values of the compressibility factor  obtained either directly from the simulations 
or indirectly from
Eq.\ (\ref{1.2}) by inserting the $\chi_{ij}$ are also included. 
\begin{table*}[tbh]
\caption{\label{table1}
Molecular dynamics values of the contact values $\chi_{ij}$ and of 
the compressibility factor $Z$ for the hard-disk binary mixtures A and B, 
with parameters given by table \protect{\ref{tab:simpara}}. 
The numbers in parentheses indicate the statistical error in the 
last digit. }
\begin{ruledtabular}
\begin{tabular}{cddddd}
\text{Mixture}&\nu&\chi_{11}&\chi_{12}&\chi_{22}&Z\\
\hline
\text{A}	&0.01&    1.013(5)    &1.015(8)     &1.02(2)   &1.0193(1)\\
\text{A}	&0.05&    1.076(4)    &1.086(3)     &1.08(2)   &1.1026(3)\\
\text{A}	&0.10&    1.162(7)    &1.181(5)     &1.20(2)   &1.2229(7)\\
\text{A}	&0.15&    1.264(5)    &1.289(6)     &1.34(1)   &1.3656(9)\\
\text{A}	&0.20&    1.383(4)    &1.426(6)     &1.48(2)   &1.537(2)\\
\text{A}	&0.25&    1.518(2)    &1.579(6)     &1.66(2)   &1.742(2)\\
\text{A}	&0.30&    1.689(3)    &1.768(8)     &1.88(2)   &1.997(2)\\
\text{A}	&0.35&    1.888(5)    &1.998(7)     &2.15(3)   &2.310(3)\\
\text{A}	&0.40&    2.132(6)    &2.28(2)      &2.51(5)   &2.708(7)\\
\text{A}	&0.45&    2.441(8)    &2.63(3)      &2.98(7)   &3.22(1)\\
\text{A}	&0.50&    2.828(5)    &3.09(2)      &3.52(4)   &3.89(1)\\
\text{A}	&0.55&    3.346(6)    &3.68(1)      &4.35(6)   &4.79(1)\\
\text{A}	&0.60&    4.03(1)     &4.51(2)      &5.41(7)   &6.05(2)\\
\text{A}	&0.65&    5.01(1)     &5.65(3)      &7.03(9)   &7.87(3)\\
\text{A}	&0.70&    6.47(2)     &7.36(7)      &9.6(2)    &10.67(5)\\
\text{A}	&0.75&    8.77(1)     &10.14(9)     &13.6(5)   &15.3(2)\\
\hline
\text{B}	&0.05&    1.076(1)    &1.084(3)     &1.09(1)    &1.1028(2)\\
\text{B}	&0.10&    1.164(1)    &1.180(3)     &1.21(1)    &1.2234(3)\\
\text{B}	&0.1482&  1.262(1)    &1.288(1)     &1.33(1)    &1.3609(5)\\
\text{B}	&0.20&    1.385(2)    &1.423(2)     &1.50(1)    &1.5376(7)\\
\text{B}	&0.25&    1.525(1)    &1.582(3)     &1.68(2)    &1.745(1)\\
\text{B}	&0.30&    1.690(1)    &1.768(2)     &1.90(1)    &1.996(2)\\
\text{B}	&0.3455&  1.873(1)    &1.978(5)     &2.15(2)    &2.282(2)\\
\text{B}	&0.3867&  2.070(3)    &2.196(6)     &2.45(2)    &2.595(3)\\
\text{B}	&0.45&    2.450(2)    &2.638(7)     &3.00(2)    &3.221(5)\\
\text{B}	&0.50&    2.843(2)    &3.096(4)     &3.59(2)    &3.891(2)\\
\text{B}	&0.55&    3.357(3)    &3.705(7)     &4.37(3)    &4.794(2)\\
\text{B}	&0.60&    4.058(2)    &4.525(6)     &5.49(2)    &6.055(3)\\
\text{B}	&0.65&    5.036(7)    &5.708(9)     &7.06(5)    &7.885(3)\\
\text{B}	&0.66&    5.281(6)    &6.00(2)      &7.46(5)    &8.348(4)\\
\text{B}	&0.67&    5.547(4)    &6.32(1)      &7.91(4)    &8.855(1)\\
\text{B}	&0.68&    5.830(5)    &6.68(1)      &8.36(4)    &9.409(1)\\
\text{B}	&0.69&    6.145(8)    &7.06(2)      &8.89(5)    &10.016(2)\\
\text{B}	&0.70&    6.498(9)    &7.45(3)      &9.55(7)    &10.684(3)\\
\text{B}	&0.75&    8.82(2)    &10.34(4)      &13.4(1)    &15.30(1)\\
\end{tabular}
\label{tab1}
\end{ruledtabular}
\end{table*}

\subsection{Contact values $\chi_{ij}$}
\begin{figure}[htbp]
\includegraphics[clip=,width=0.90 \columnwidth]{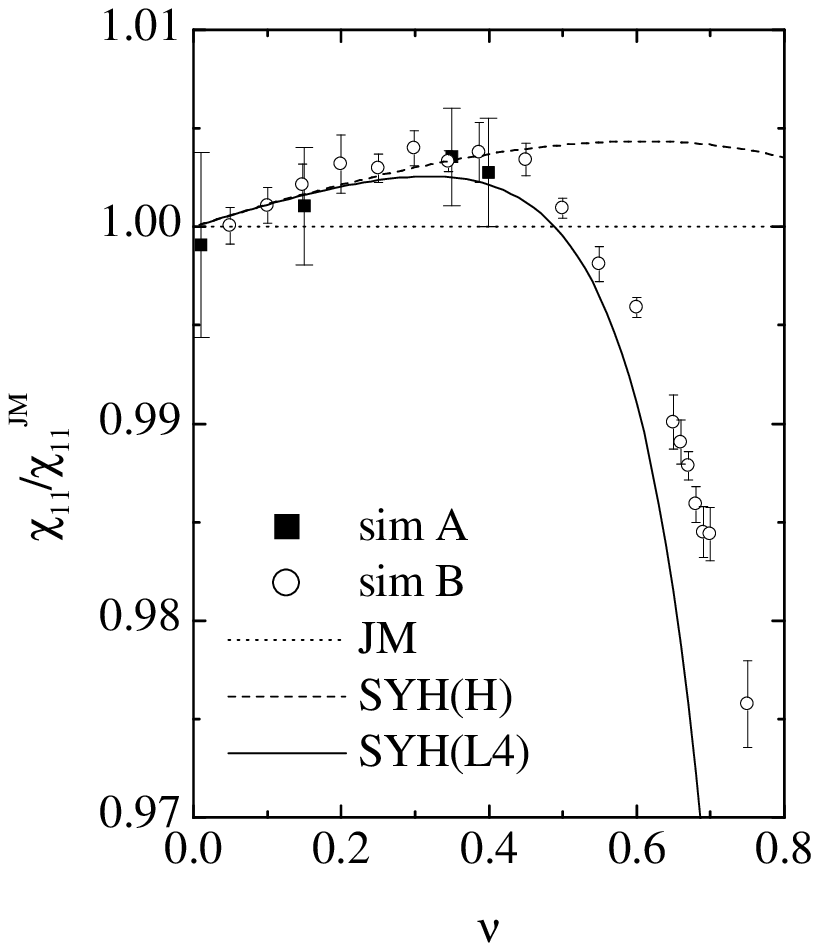}
\caption{Density dependence of the ratio $\chi_{11}/\chi_{11}^{\text{JM}}$ 
for a hard-disk binary mixture with parameters as given
in Table \protect{\ref{tab:simpara}}.  
The filled squares and open circles  represent our simulation 
data A and B, respectively. 
The lines represent the ratio $\chi_{11}^{\text{SYH}}/\chi_{11}^{\text{JM}}$ 
for case B
with $\chi(\nu)=\chi^{\text{H}}(\nu)$ (dashed line) and 
with $\chi(\nu)=\chi^{\text{L4}}(\nu)$ (solid line); the dotted
line represents Eq.\ (\ref{1.5}). 
The theoretical curves for case A are practically indistinguishable from those for 
case B and so they are not plotted.\label{g11}}
\end{figure}

\begin{figure}[htbp]
\includegraphics[clip=,width=0.90 \columnwidth]{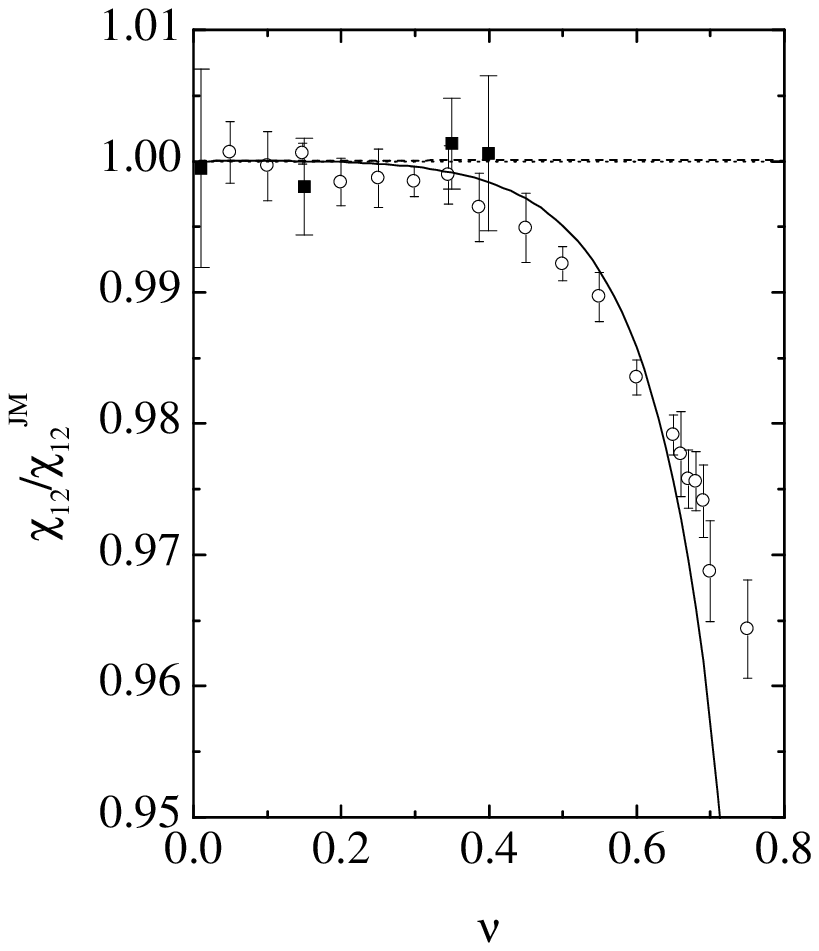}
\caption{Same as in Fig.\ \protect\ref{g11}, 
except for $\chi_{12}/\chi_{12}^{\text{JM}}$. \label{g12}}
\end{figure}

\begin{figure}[htbp]
\includegraphics[clip=,width=0.90 \columnwidth]{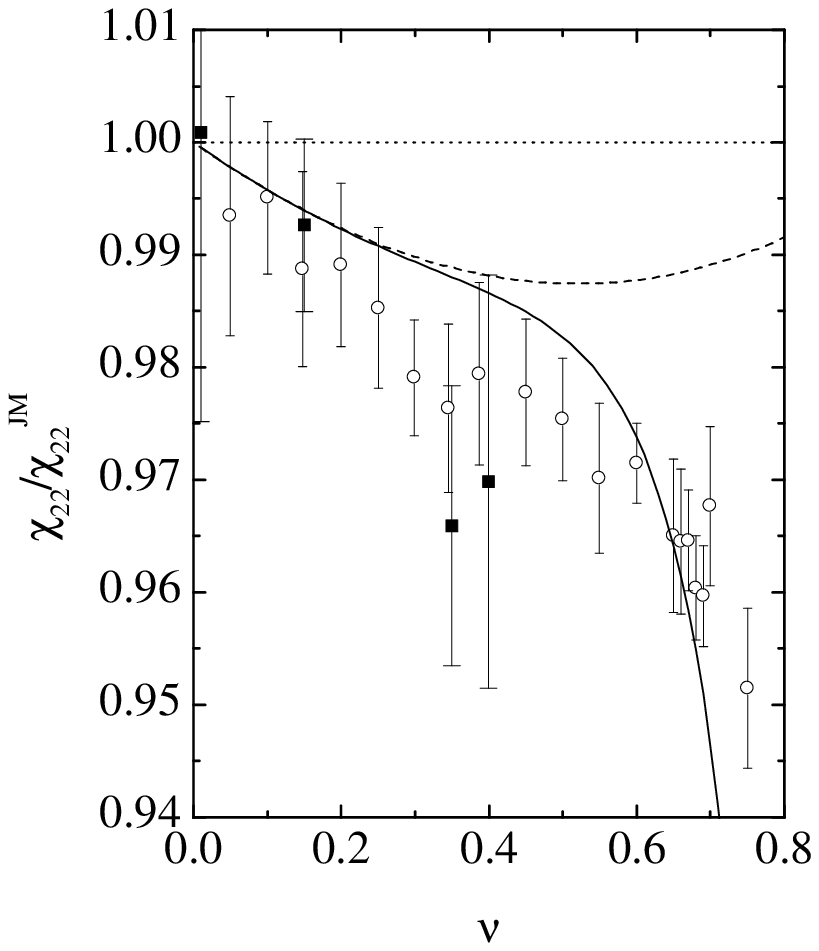}
\caption{Same as in Fig.\ \protect\ref{g11}, 
except for $\chi_{22}/\chi_{22}^{\text{JM}}$. \label{g22}}
\end{figure}

In Figs.\ \ref{g11}, \ref{g12}, and \ref{g22} we have plotted the 
simulation values for the ratios 
$\chi_{11}/\chi_{11}^{\text{JM}}$, $\chi_{12}/\chi_{12}^{\text{JM}}$, 
and $\chi_{22}/\chi_{22}^{\text{JM}}$, respectively, 
against the area fraction $\nu$. 
The data corresponding to case A are shown only for the densities 
not considered in case B, namely $\nu=0.01$, 0.15, 0.35, and 0.40.
The ratio $\chi_{ij}/\chi_{ij}^{\text{JM}}$ represents a ``quality factor'' 
of the simulation data with respect to the JM approximation 
(\ref{1.5}). 
Figures  \ref{g11}--\ref{g22} also include the ratios 
$\chi_{ij}^{\text{SYH}}/\chi_{ij}^{\text{JM}}$, where in Eq.\ (\ref{1.12}) we 
have taken  $\chi(\nu)=\chi^{\text{H}}(\nu)$ [cf.\ Eq.\ (\ref{1.7})] and 
$\chi(\nu)=\chi^{\text{L4}}(\nu)$ [cf.\ Eq.\ (\ref{1.9})] 
for the monodisperse fluid. 

\subsubsection{Low densities}

We observe that $\chi_{ij}^{\text{SYH}}/\chi_{ij}^{\text{JM}}$ with  
both prescriptions $\chi(\nu)=\chi^{\text{H}}(\nu)$ and
 $\chi(\nu)=\chi^{\text{L4}}(\nu)$   are practically indistinguishable 
up to $\nu\approx 0.3$. This is consistent with the fact that in 
that domain of low and moderate densities the correction (\ref{1.9}) to 
Henderson's EOS is irrelevant, as shown in Fig.\ \ref{gmono}. 
On the other hand, some limitations of Eq.\ (\ref{1.5}) are already 
apparent in the range $0<\nu\lesssim 0.3$: the JM approximation  
(slightly) underestimates the small-small contact value 
($\chi_{11}^{\text{JM}}<\chi_{11}$), while it overestimates the  large-large 
contact value ($\chi_{22}^{\text{JM}}>\chi_{22}$). 
These two effects, which are not linked to the use of  $\chi^{\text{H}}(\nu)$, 
are reasonably well captured by Eq.\ (\ref{1.12}), 
as expected from (\ref{1.14}). 
In the case of the cross contact value, we have $\chi_{12}\simeq 
\chi_{12}^{\text{JM}}$ for $0<\nu\lesssim 0.3$, in agreement 
with the fact that $z_{12}\simeq 1$ in our mixtures A and B.

\subsubsection{High densities}
In the high-density domain $\nu\gtrsim 0.3$,  the simulation data 
clearly deviate from both  Eq.\ (\ref{1.5}) and Eq.\ (\ref{1.12}) 
when the latter is combined with $\chi(\nu)=\chi^{\text{H}}(\nu)$. 
Both theories tend to overestimate the contact values, what  is 
essentially a trait inherited from Henderson's EOS. On the other hand, 
a much better agreement is obtained when  Eq.\ (\ref{1.12}) is 
combined with Eq.\ (\ref{1.9}) for the monodisperse fluid. The remaining 
deviations of the latter theory from the simulation values are 
a reflection of the approximate character of Eq.\ (\ref{1.12}) rather 
than that of Eq.\ (\ref{1.9}), in view of Fig.\ \ref{gmono}. 
Part of the deviations for the highest densities may be due to the 
proximity to crystallization. In the monodisperse case, it is known 
that the hard-disk fluid undergoes a freezing transition (possibly 
mediated by a hexatic phase\cite{J99}) at an area fraction 
$\nu\simeq 0.7$.\cite{J99,Lu01,Lu01b,Lu04,Lu02,kawamura79} In any case, 
polydispersity tends to increase the freezing transition 
density. 
{For mixture A, our simulations indicate that the transition 
takes place between $\nu=0.75$ and $\nu=0.8$. For more details on very high densities
in the mono- and bi-disperse situations see Refs.\ \onlinecite{Lu01} 
and \onlinecite{Lu02}, respectively. }

\subsection{Equation of State}

\begin{figure}[htbp]
\includegraphics[clip=,width=0.90 \columnwidth]{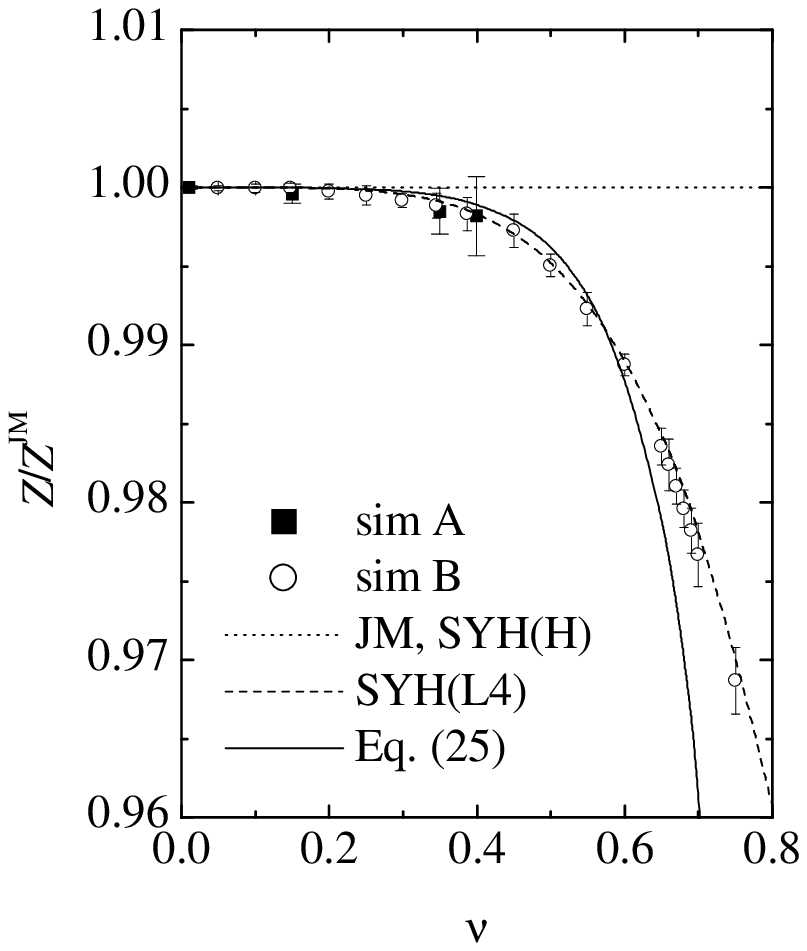}
\caption{
Density dependence of the ratio $Z/Z^{\text{JM}}$ for a
hard-disk binary mixture with parameters as given in Table
\protect{\ref{tab:simpara}}.
The filled squares  and open circles  represent our simulation 
data A and B, respectively. 
The lines represent the ratio $Z^{\text{SYH}}/Z^{\text{JM}}$ for case B with
$\chi(\nu)=\chi^{\text{H}}(\nu)$ (dotted line -- unity) and
$\chi(\nu)=\chi^{\text{L4}}(\nu)$ (solid line).  The dashed line
is the empirical relation (\protect\ref{empirical}) proposed in Ref.\ 
\protect{\onlinecite{Lu01b}}.
\label{Z}}
\end{figure}

Although in this paper we have been mainly concerned with the contact 
values of the RDFs, it is worth considering the compressibility 
factor $Z\equiv p/\rho k_BT$. The simulation values of the ratio $Z/Z^{\text{JM}}$, 
where $Z^{\text{JM}}$ is given by Eq.\ (\ref{1.10}) are plotted in 
Fig.\ \ref{Z}. We observe that the JM EOS is fairly 
accurate for $\nu\lesssim 0.4$, 
even though the individual values $\chi_{ij}^{\text{JM}}$ are not that 
good in the same density range. This is mainly due to a fortunate 
``cancellation of errors'' ($\chi_{11}^{\text{JM}}<\chi_{11}$, 
while $\chi_{22}^{\text{JM}}>\chi_{22}$). 
As a matter of fact, as already mentioned in Sec.\ \ref{sec1bis}, 
the recipe (\ref{1.13}) in combination with $\chi(\nu)=\chi^{\text{H}}(\nu)$ 
becomes identical with $Z^{\text{JM}}$. 
Nevertheless, the JM approximation again overestimates the simulation 
data for higher densities ($\nu\gtrsim 0.4$). 
When $\chi(\nu)=\chi^{\text{L4}}(\nu)$ is used, Eq.\ (\ref{1.13}) becomes 
quite reasonable, although it slightly underestimates the mixture pressure of 
the fluid at the highest densities.

Interestingly, a recently proposed empirical correction to the
EOS 
\beq
Z = 1+(1-a\nu^4)\left(Z^{\text{JM}}-1\right), \quad a=0.1,
\label{empirical}
\eeq
see Eq.\ (20) in Ref.\ \onlinecite{Lu01b}, works also pretty well
for the parameter set used here, however, without theoretical foundation. 
The excellent performance of Eq.\ (\ref{empirical}) for $x_1\simeq 0.8$ 
and $\sigma_1/\sigma_2=\frac{1}{2}$ does not necessarily 
extend to other compositions. In fact, Eq.\ (\ref{empirical}) is not as 
good as Eq.\ (\ref{1.9}) in the monodisperse case.

\section{Conclusion\label{sec4}}

In this paper we have presented  
molecular dynamics results for the contact values 
$\chi_{ij}=g_{ij}(\sigma_{ij})$ of the radial distribution functions
(RDFs) $g_{ij}(r)$ of binary mixtures of additive hard disks. 
As a representative case we have fixed mole fractions $x_1\simeq 0.8$ 
and $x_2\simeq 0.2$, and used a diameter ratio 
$\sigma_1/\sigma_2=\frac{1}{2}$.
A set of numerical values of the area fraction $\nu$ have been considered, 
covering the dilute, the intermediate, and the dense regime.

The simulation results have been used to assess the reliability of 
theoretical expressions previously proposed in the literature. 
Until recently, practically the only proposal was the one by 
Jenkins and Mancini,\cite{JM87} which is expressed by 
Eq.\ (\ref{1.5}). This approximation succeeds in capturing the main 
trends in the intricate dependence of $\chi_{ij}$ on the area fraction 
and the composition of the mixture, even at a quantitative level. 

On the other hand, our simulation data expose some (small) limitations 
of $\chi_{ij}^{\text{JM}}$: already for low and moderate densities 
($\nu\lesssim 0.3$) the JM recipe tends to underestimate the 
small-small correlation value and overestimate the large-large value; 
for higher densities ($\nu\gtrsim 0.3$), 
$\chi_{ij}^{\text{JM}}$ overestimates the correct values, 
an effect that can be traced back to the fact that 
$\chi_{ij}^{\text{JM}}$ is strongly tied to Henderson's 
EOS.\cite{H75}

These two shortcomings are widely corrected by a recent proposal made 
by Santos et al.,\cite{SYH02} Eq.\ (\ref{1.12}). While 
$\chi_{ij}^{\text{JM}}$ is a linear function of the parameter 
$z_{ij}$ defined by Eq.\ (\ref{1.6}), 
$\chi_{ij}^{\text{SYH}}$ is a quadratic function. 
This higher order approach allows $\chi_{ij}^{\text{SYH}}$ 
to satisfy an extra consistency condition in the limit of highly 
asymmetric mixtures.\cite{SYH02} 
Moreover, $\chi_{ij}^{\text{SYH}}$ can be used in conjunction with any 
desired expression for the contact value of the monodisperse fluid. 
When instead of Henderson's expression the one recently proposed by 
one of us\cite{Lu01,Lu02} is employed, 
$\chi_{ij}^{\text{SYH}}$ exhibits a reasonable agreement with the 
simulation data for $\nu\gtrsim 0.3$. 

In spite of this, it can be observed that $\chi_{ij}^{\text{SYH}}$ tends 
to underestimate the simulation data for very high densities 
($\nu\gtrsim 0.6$), so that an even better approximation is needed 
in that extreme, high density fluid regime. {}From this point of view,
we hope that our simulation data will be helpful to test the 
accuracy of other future theoretical proposals that have been or will 
be made.\\
~\\

\acknowledgments
S.L. acknowledges helpful discussions with J.T. Jenkins and M. Alam,
as well as the financial support of the DFG (Deutsche Forschungsgemeinschaft,
Germany) and FOM (Stichting Fundamenteel Onderzoek der Materie, 
The Netherlands) as financially supported by NWO (Nederlandse Organisatie
voor Wetenschappelijk Onderzoek).
A.S. is grateful to M. L\'opez de Haro for discussions about the  
topic of this paper. The research of A.S. has been partially supported 
by the Ministerio de Ciencia y Tecnolog\'{\i}a (Spain) through grant 
No.\ FIS2004-01399 and  by the European
Community's Human Potential Programme under contract HPRN-CT-2002-00307,
DYGLAGEMEM.

\end{document}